# Electrically and Magnetically Tunable Charge-Density-Wave Transport in Quasi-2D *h*-BN/1*T*-TaS₂ Thin-Film Heterostructures

Jonas O. Brown[1,2], Maedeh Taheri[1,2], Nicholas R. Sesing[3], Tina T. Salguero[3], and Alexander A. Balandin[1,2,4*]

[1]Phonon Optimized Engineered Materials Laboratory, Department of Materials Science and Engineering, University of California, Los Angeles, California 90095 USA

[2]California NanoSystems Institute, University of California, Los Angeles, California 90095 USA

[3]Department of Chemistry, University of Georgia, Athens, Georgia 30602 USA

[4]Center for Quantum Science and Engineering, University of California, Los Angeles, California 90095 USA

* Corresponding author (A.A.B.): balandin@seas.ucla.edu; website: https://balandin-group.ucla.edu/

# Abstract

Controlling collective electronic phases in low-dimensional materials is a central challenge for developing technologies based on charge-density waves. Here, we report how perpendicular electric and magnetic fields can be utilized to tune the charge-density-wave transport in the quasi-two-dimensional material $1T$-TaS$_2$. Using $h$-BN-encapsulated thin-film heterostructures with both top-gate and bottom-gate configurations, we find that electrical gating produces a non-monotonic shift in the depinning threshold—behavior distinct from quasi-one-dimensional charge-density-wave systems. We further show that a perpendicular magnetic field increases the threshold voltage for domain depinning and can drive the nearly commensurate-to-incommensurate charge-density-wave phase transition, demonstrating magnetic control over a two-dimensional electron–lattice condensate. The obtained results shed light on mechanisms governing charge-density-wave domain dynamics and reveal combined electrical and magnetic field control as a strategy for engineering low-power-dissipation devices and electronics for extreme environments.

**Introduction**

Charge-density-wave (CDW) materials offer a promising platform for low-power and high-frequency electronic devices because their collective electron–lattice states can be reconfigured by external stimuli. The threshold field for CDW depinning and the ability to modulate phase transitions are, therefore, key parameters for device design. The CDW phase of some metals with a strongly anisotropic Fermi surface constitutes a periodic modulation of electronic charge density and a corresponding lattice distortion[1,2]. Early investigations of Peierls-type CDWs in bulk quasi-one-dimensional (1D) metallic crystals demonstrated multiple resistive phases, non-Ohmic electron transport, and voltage-driven current oscillations[3,4,5,6]. Although quasi-one-dimensional (1D) Peierls-type CDW crystals exhibit well-understood depinning phenomena, the influence of electric and magnetic fields on CDW dynamics in two-dimensional (2D) systems is far less explored. The gate-field tuning of the depinning threshold, $E_T$, has been demonstrated in quasi-1D NbSe$_3$ and TaS$_3$[7] electronic materials. However, the effect of external fields on CDW *domains* in quasi-2D materials, where coherent sliding is absent, remains limited.

The physics underlying CDW formation and behavior varies significantly across different crystal structures and dimensionalities. In quasi-1D systems, the Peierls instability dominates[1,8-12], whereas layered quasi-2D van der Waals materials introduce more intricate interplay between Fermi surface nesting and electron-phonon coupling[13,14]. The quasi-2D material 1$T$-TaS$_2$ provides an interesting platform for this investigation due to its rich sequence of CDW transitions: a commensurate CDW (C-CDW) phase below *216 K*, a nearly commensurate CDW phase (NC-CDW) between *216 K* and *352 K*, and an incommensurate CDW (IC-CDW) phase above *352 K*[15-22]. In the NC-CDW phase, transport is governed by nanoscale C-CDW domains separated by metallic IC-CDW domain walls, producing a complex interplay of correlation effects, percolation, and electron–phonon coupling. Prior studies have shown that large in-plane electric fields do not induce coherent CDW sliding in 1$T$-TaS$_2$; instead, transitions and hysteresis often arise from local Joule heating[23-27]. Still, CDW domain depinning does occur and manifests as fluctuation features in $dI/dV$ and low-frequency noise[19-21]. We have previously demonstrated that a perpendicular electric field can modulate and even induce the NC-CDW ↔ IC-CDW transition, underscoring the sensitivity of the CDW domain network to gating[16]. However, no comprehensive studies have addressed how perpendicular electric or magnetic fields influence the threshold for domain depinning, likely due to difficulties of monitoring the threshold field, $E_T$, without a clear onset of collective current.

Understanding these field effects is not only of fundamental interest but also technologically relevant. Neuromorphic devices that leverage the NC-CDW $\leftrightarrow$ IC-CDW transition utilize oscillatory phase dynamics to evolve towards solutions for benchmark optimization problems [28,29]. Frequencies may be able to approach the *THz* range due to the ultrafast switching of NC-CDW domain reconfiguration[30-32]. Nanoscale $1T$-TaS$_2$ devices have demonstrated that Joule heat-driven transitions can reach *GHz* frequencies[26]. Because dimensionality[33], disorder[34,35], and correlation effects[36] critically shape CDW behavior in 2D materials, clarifying how external fields modulate the depinning threshold and phase stability is essential for advancing CDW-based electronic components.

Electric and magnetic fields perturb both the single-particle electronic structure and the collective CDW condensate[3,9]. In CDW systems, such perturbations can alter pinning potentials, modify domain-wall connectivity, or stabilize competing phases[7,16,37,38]. In $1T$-TaS$_2$, these interactions are further complicated by strong correlations[20] and a polaronic commensurate superstructure[39-44], making field-driven effects fundamentally distinct from those in quasi-1D Peierls systems[23-25]. Gate fields can redistribute carriers within the metallic domain walls[16]. Whereas magnetic fields may influence Fermi-surface geometry or induce Anderson-Mott type localization effects, thereby shifting depinning thresholds or CDW phase boundaries[37,38].

Motivated by these open questions, we have investigated how perpendicular electric and magnetic fields influence CDW domain depinning and phase transitions in $h$-BN/$1T$-TaS$_2$ heterostructures. By utilizing both top-gate and bottom-gate architectures, we examined possible differences due to the penetration depth of the electric field. We studied the strength, reversibility, and monotonic nature of magnetic field-induced changes to device characteristics to determine whether a magnetic field alone can trigger the NC-CDW $\rightarrow$ IC-CDW transition. The results and insights presented here help advance the understanding of field-tunable CDW dynamics in quasi-2D materials and establish a foundation for practical electronic applications based on $1T$-TaS$_2$.

## 1. Results and Discussion

### 2.1. Device Fabrication and Current-Voltage Characteristics

Bulk $1T$-TaS$_2$ single crystals were synthesized using chemical vapor transport (CVT) following established methods[46-49]. Mechanically exfoliated thin films (*10 nm - 100 nm* thicknesses) from CVT-grown $1T$-TaS$_2$ crystals were transferred onto Si/SiO$_2$ substrates. The film thickness was confirmed via atomic force microscopy (AFM) and optical contrast analysis. To protect the $1T$-TaS$_2$ crystal from oxidation and surface contamination, hexagonal boron nitride ($h$-BN) was used as a capping layer. The source-drain and top gate contacts were patterned using electron beam lithography, followed by atomic layer etching to remove the $h$-BN layer underneath the source-drain contacts. The contacts were deposited with electron beam evaporation of Ti / Au with *10 nm / 100 nm* thicknesses and with lateral dimensions of *1 μm — 3 μm*. Two types of gating were used in this study, top gating through the ~*30 nm h*-BN dielectric and back gating through the *300 nm* SiO$_2$ substrate. The back gate consisted of a silver contact deposited on the highly doped (p$^+$) Si substrate in a region of the substrate where the SiO$_2$ was removed. Over 20 different $h$-BN/$1T$-TaS$_2$ devices with varying channel thickness and source-drain separation, $L < 3$ *μm*, have been fabricated and tested in this study. The data presented in the main text, below, was obtained with device 1 (channel thickness ~*30 nm*) for top gating experiments, device 2 (~*15 nm*) for bottom gating experiments, device 3 (~*100 nm*) for magnetoresistance experiments, device 4 (~*10 nm*) for temperature driven phase transitions under a perpendicular magnetic field, and device 5 (~*15 nm*) for the current driven phase transitions under a magnetic field. In Figure 1 (a), we present a schematic of $h$-BN-encapsulated top-gate $1T$-TaS$_2$ devices. An AFM image and height profile of a channel region of a representative $h$-BN/$1T$-TaS$_2$ CDW device are shown in Figures 1 (b) and (c), respectively. Optical microscopy images of the tested devices can be found in the Supplemental Information.

[**Figure 1: Device structure and electrical characteristics.** (a) Schematic illustration of a $h$-BN/$1T$-TaS$_2$ heterostructure device with a top gate. (b) Tilted AFM image of a representative device channel region with metal electrodes on top. (c) The AFM height profile showing the thickness of the $h$-BN capping layer (~*33 nm*), the $1T$-TaS$_2$ channel (~*40 nm*), and the metal contact. (d) Typical *I-V* characteristics of a $1T$-TaS$_2$ device measured at RT. The arrows indicate the direction of the source-drain bias sweep; $V_H$ represents the NC→IC-CDW phase transition, while $V_L$ represents the IC→NC-CDW phase transition. The bias voltage indicated by $V_D$ corresponds to the onset of domain depinning, determined from the derivative $I$-$V$s. (e) Magnified image of the hysteresis region of the $I$-$V$s, showing the effect of the gate. (f) Derivative, $dI/dV$, characteristics of the device showing the threshold for the CDW domain depinning as $V_D^f$ for the forward bias and $V_D^r$ for the reverse bias. The voltages which correspond to the hysteresis, $V_H$ and $V_L$, are also shown. The data in panels (d), (e), and (f) are for device 1.]

A typical current-voltage ($I$-$V$) characteristic for a 1$T$-TaS$_2$ CDW device, measured at RT, is shown in Figure 1 (d). The channel thickness for this device is ~*10 nm*. The direction of the voltage sweep is indicated by the curved arrows. The hysteresis results from thermally driven transitions between the NC→IC-CDW phases at a bias voltage of $V_H$ and the reverse IC→NC-CDW phase transition at a bias of $V_L$. The local heating is produced by the source-drain current in the nanoscale-thick channel. The NC-CDW phase is more resistive than the IC-CDW phase. The low-bias, temperature-dependent resistivity measurements show the NC→IC-CDW transition onset at ~*365 K* on heating and the IC→NC-CDW transition at ~*340 K* on cooling (see the Supplemental Information). This behavior is consistent with prior studies and theoretical models of the CDW order parameter[16-21].

Figure 1 (e) presents the gate voltage modulation effects on the field-driven phase transitions $V_H$ and $V_L$ of a top-gated device at RT. The results show that applying a top gate bias (-*5 mV* to *2 mV*) shifts the CDW phase transition. The observed partial reversibility suggests that the gate electric field induces permanent modifications to the C-CDW domain structure, potentially through charging by the surrounding IC-CDW phase, shifting, or merging—resulting in shifted $I$–$V$ hysteresis loops[16]. This behavior contrasts with quasi-1D CDW materials, such as NbSe$_3$, where gate-induced variations cause reversible modulation of the depinning threshold[7]. Owing to the high concentration of carriers in 1$T$-TaS$_2$, the estimated penetration depth is a few *nm*. Gating in quasi-2D 1$T$-TaS$_2$ likely affects the C-CDW domains near the gate dielectric, where the gate shifts the CDW phase transition non-monotonically[16]. Since the thickness of the channel layer is larger than the electric field penetration depth and because of the absence of a collective CDW current component in this 2D material, the modulation of the hysteresis is small[16]. The irreversibility indicates that the gate electric field induces permanent modifications to the C-CDW domains. Under sufficiently strong fields, the domains may become charged by electrons from the surrounding IC-CDW phase, undergo positional shifts, or merge with adjacent domains, resulting in structural changes. During the measurements, the control of thermal effects was ensured by spacing *I-V* sweeps with two-minute intervals. The control measurements confirmed that the hysteresis loop area remains nearly constant, indicating the observed shifts of the hysteresis by the gate bias originate from direct electric field influence rather than local heating, which typically reduces hysteresis size[23,28]. The source-drain current softens the CDW domains and induces the hysteresis, while the electrical field created by the gate modifies the domain structure and charge.

To monitor the onset of domain depinning in 1$T$-TaS$_2$, we use derivative *I-V* characteristics. Figure 1 (f) shows *dI/dV vs.* applied bias voltage for forward and reverse bias at RT. The derivative *I-V*s are utilized following the methodology reported by us earlier in Refs. [16, 23, 28]. Domain depinning in 2D CDW materials is different from that in quasi-1D CDW materials. The quasi-1D CDW materials show an abrupt increase in current at the depinning threshold due to the onset of CDW sliding and emergence of the collective current[6]. The domain depinning in quasi-2D systems like 1$T$-TaS$_2$ does not result in noticeable collective current and does not change the linear character of the *I-V* curves[23,28]. After depinning, the C-CDW domains become more susceptible to changes induced by the external fields. The onset of domain depinning is shown as $V_D$ in Figure 1 (f). While not visible in *I-V*s, the domain depinning in 1$T$-TaS$_2$ is clearly seen as the onset of spikes in the derivative characteristics, $dI/dV$, and in low-frequency noise[25]. In Figure 1 (f), the threshold for domain depinning, $V_D$, is denoted as $V_D^f$ for the forward bias and $V_D^r$ for the reverse bias. The phase transition points, $V_H$ and $V_L$, which correspond to the width of the hysteresis, are also shown at higher voltages. Note that $V_H$ and $V_L$ in Figure 1 (f) coincide exactly with the hysteresis borders in Figure 1 (d). One can observe the phase transition in both *I-V*s and derivative characteristics, *dI/dV*, but the CDW domain depinning is only visible in the derivative characteristics as the onset of spikes.

## 2.2. Top-Gate Field Effect on CDW Domains

The effect of the applied gate bias, $V_G$, on the domain depinning threshold at RT is presented in Figure 2 (a) for the same top-gated device as in Figure 1. Tracing the onset of the current fluctuations, *i.e.* appearance of spikes in derivative *I-V*s, for different applied gate bias reveals a non-monotonic dependence of the depinning threshold voltage for both the forward and reverse source-drain biases (see Figure 2 (b)). The gate bias is applied, starting from $V_G = $ *-5 mV* to *3 mV*, with *1 mV* steps. This corresponds to a field of $E_G=$ *-0.17 mV/nm* to $0.1\ mV/nm$ which is comparable to the horizontal channel field required to depin the CDW domains, $E_D = $ *0.1 mV/nm* to *0.18 mV/nm*. Uncertainty is represented by the red (forward bias) and violet (reverse bias) shadowed regions determined using B-spline interpolation of the standard deviation obtained from four control measurements at $E_G = $ *0 V* for both $E_D$ and $E_H$. The uncertainty is calculated as the standard deviation from the mean of four consecutive source-drain sweeps conducted without a gate. Notice the approximately even symmetry of the depinning threshold with respect to the gate bias.

The gate effect on the depinning threshold, $V_D$, and on the onset of the hysteresis, $V_H$, associated with the NC-CDW – IC-CDW transition can also be quantified by the voltage modulation due to the gate bias. We define the voltage modulation due to the applied gate bias as $\Delta V/V = \left( V_{H,D}(V_G) - V_{H,D}(0) \right) / V_{H,D}(0)$ where $V_{H,D}(V_G)$ represents the threshold voltage for the hysteresis or depinning at a certain $V_G$ value while $V_{H,D}$ *(0)* represents $V_{H,D}$ at $V_G = 0\ V$. The gate voltage modulations for the depinning and phase transition are shown in Figures 2 (c) and (d), respectively. It should be noted that the voltage modulation of the depinning onset $V_D$ is significantly larger than the voltage modulation of the phase transition $V_H$. Additionally, the CDW domain depinning modulation exhibits a non-monotonic dependence on the applied gate, with approximately even symmetry, while the phase transition modulation decreases monotonically with an increasing gate field in the top-gated device in the low-field regime $E_G \sim E_D$.

**[Figure 2: Electrical gating of the charge-density-wave domain depinning and hysteresis**. (a) *dI/dV* as a function of the source-drain bias during forward sweep at gate biases from *-5 mV* to *3* *mV*. (b) The onset of domain depinning as a function of gate bias for forward (red) and reverse (violet) source-drain bias. (c) The voltage modulation, $\Delta V/V$, as a function of the gate electric field for the depinning onset $V_D$ in the forward sweep. (d) The voltage modulation for the NC→IC transition $V_H$. The electric field created by the top gate is of the same order of magnitude as the source-drain bias at the onset of the CDW domain depinning. The data are presented for device 1.]

### 2.3. Bottom-Gate Field Effect on CDW Domains

Since the electric field penetration depth is limited in the CDW material, there is a possibility that the gate effect in the top-gate and bottom-gate devices will be different. The bottom gate extends over a larger area than the top gate and can be used to supply a higher bias voltage. The gate dielectrics in the two designs are also different: *h*-BN for the top gate (see Figure 1 (a)) and $SiO_2$ for the bottom gate (Figure 3 (a)). In the back-gated structure, the thickness of the $SiO_2$ gate dielectric is *300 nm*. The back-gate design was used to observe the effect of the gate on CDW domain depinning and phase transitions at higher fields, uniformly distributed along the entire channel. A maximum field of *±0.2 V/nm* was achieved for back-gated devices compared to a maximum field of *±0.17 mV/nm* for the top-gated device. The voltage modulation of the depinning onset $\Delta V_D/V_D$ with respect to the gate field, $V_G$, is shown in Figures 3 (b) and (c) for *200 K* and *300 K*, respectively. In these measurements, the cross-plane electric field created by the bottom gate is larger than the electric field created by the source-drain bias voltage along the device channel, *i.e.*, $E_G > 100E_D$. In this high-field regime, $\Delta V_D/V_D$ scales non-monotonically with $V_G$ across all temperatures studied. For comparison, Figure 3 (d) shows the back-gated device in the low-field regime, *i.e.*, $E_G \sim E_D$. Like in the top-gated devices, the voltage modulation of $V_D$ is approximately even in the gate voltage. The observed dependencies for the CDW domain depinning in quasi-2D 1*T*-TaS$_2$ differ from those reported for gating of quasi-1D materials such as NbSe$_3$, TaS$_3$, and NbSe$_3$, where the voltage modulation of the depinning threshold changes monotonically with the gate field[7].

In quasi-1D CDW materials, one distinguishes the strong and weak CDW pinning mechanisms [1]. In the strong pinning scenario, impurities exert a significant potential, locally anchoring the CDW phase and disrupting the CDW phase coherence over short lengths. The depinning threshold field scales linearly with impurity concentration ($E_D \propto n_i$)[9]. In the weak pinning regime, the impurity potential is relatively weak, allowing the CDW to maintain a degree of phase coherence over extended regions. The depinning threshold field scales with the square of impurity concentration ($E_D \propto n_i^2$)[9]. It has been reported for $1T$-TaS$_2$ that, based on the time-averaged scattering measurements, this material follows the weak pinning scenario[18], similar to many quasi-1D CDW materials[4,7]. A possible microscopic mechanism of the weak pinning is related to the formation of dislocations by the applied gate bias[7,52-54]. The dislocation density, proportional to $E_D$, results in even symmetry of the depinning threshold to the gate voltage. At low gate electric fields, close in magnitude to the in-plane electric field at $V_D$, there is even symmetry of $E_D$ with respect to $V_G$ (see Figure 2 (b)). Thus, the gate field causing dislocations is a likely scenario. However, at stronger gate fields, the even symmetry is lost, suggesting the dislocation-based explanation may not fully account for the behavior in the high-field regime. Other mechanisms can include a CDW gap modulation in the C-CDW domains or asymmetric strain effects from dielectric interfaces[7,52-54].

[Figure 3: Electrical gating of the charge-density-wave domains at various temperatures. (a) Schematic illustration of a $h$-BN/$1T$-TaS$_2$ heterostructure device with a bottom gate. (b) The voltage modulation of the depinning onset $\Delta V_D/V_D$ in the high-field regime $E_G > 100E_D$ at $T = 300$ $K$ (c), at $T = 200 K$ (d), and in the low-field regime $E_G \sim E_D$ at $T = 200 K$. The data are presented for device 2.]

Comparing gate-modulated CDW domain depinning in quasi-2D $1T$-TaS$_2$ with the gate-modulated depinning of the coherent CDW in quasi-1D materials such as NbSe$_3$ and TaS$_3$ reveals important differences. The threshold field for CDW domain depinning in our thin-film $1T$-TaS$_2$ devices ($E_D \sim 1$ V/$\mu m$ — 2 V/$\mu m$, $H < 35$ nm) is higher than that for CDW depinning and sliding in thick NbSe$_3$ and TaS$_3$ crystals ($E_D \sim 0.035$ V/$\mu m$ — $0.045$ V/$\mu m$, $H <1$ $\mu m$)[7]. This is consistent with the expected inverse scaling with layer number in the weak pinning regime[3]. The gate-field to channel-field ratios required to modulate the depinning threshold to a similar magnitude, e.g., $\Delta V_D \sim 30\%$ — $40\%$, differ significantly. Quasi-1D devices with SiO$_2$ bottom gates ($55$ nm) require $E_G/E_D \sim 10^4$ at cryogenic temperatures ($30 K$ — $145 K$),[7] while $h$-BN top-gated ($30$ nm) and SiO$_2$ bottom-gated (*300 nm*) 1*T*-TaS$_2$ achieve comparable modulation with $E_G/E_D \sim 0.1 - 0.2$ at RT. Another difference is that quasi-1D devices require sufficiently large channel widths (*100 µm — 800 µm*) to achieve a noticeable gate modulation. In quasi-2D 1*T*-TaS$_2$, the modulation of $V_D$ was achieved in small-channel-width devices (*1 µm — 2 µm*). Together, these factors make 2D CDW materials like 1*T*-TaS$_2$ promising for threshold field modulation.

### 2.4. Magnetic Field Effect on Charge Density Wave Domains

Characterizing both electrical and magnetic responses is important for defining the functionality of novel electronic materials. At low temperature, 1*T*-TaS$_2$ transitions to the C-CDW phase and undergoes a structural reconstruction into 13-Ta clusters, generating a narrow-gap insulating state stabilized by specific interlayer stacking orders[20]. In this regime, the application of a magnetic field is expected to induce noticeable changes in transport through delocalizing spin-polarized electrons[40,45] and wavefunction shrinkage and interference[55,56]. Conduction is governed by the variable-range hopping (VRH) in the C-CDW phase between localized states arising from stacking faults and domain boundaries[53-57]. This is confirmed by the $R(T) \propto exp(T^{-1/3})$ dependence in the C-CDW phase for both the 100 nm device and the bulk device (see Supplemental Information), matching previous reports[53]. Figure 4 (a) shows the magnetoresistance, $MR = (\rho(B)-\rho(0))/\rho(0) = \Delta\rho/\rho(0)$, of a 1*T*-TaS$_2$ channel with a thickness of *~100 nm*. A finite, even-in-field magnetoresistance appears below *~50 K* and reaches a maximum of *~2.7%* at $B = \pm9$ *T* and $T = 1.8$ *K*. The magnetoresistance scales as $B^2$ up to *~4 T* (fitting shown in Figure S7 (c)), then grows linearly at higher fields. While bulk crystals exhibit magnetoresistance as large as *~20%* when $B \sim 7.5$ *T* at $T = 1.4$ *K*[44] and , *>35%* when $B = 9$ T at $T = 1.8K$ (see the Supplemental Information), 100 *nm* thick films show values about an order of magnitude smaller. The magnitude of the MR in 1*T*-TaS$_2$ is dependent on the crystal quality, stacking order, and long-range phase coherence. In thin films, the reduced volume limits stacking disorder pathways, and the electron transport can become dominated by surface states[39,554]. The magnetoresistance is positive and follows a quadratic dependence, conventionally described by the VRH relation $ln[\rho(B)/\rho(0)]=\alpha B^2$[54-56]. The coefficient $\alpha$ depends on the specific hopping model[54].

We now turn to the effect of the magnetic field on the resistance *vs* temperature hysteresis induced by the NC-CDW—IC-CDW phase transition. Although the single particle resistivity of 1$T$-TaS$_2$ is not modified by the magnetic field at high temperatures, the CDW transition temperature is affected. Figure 4 (b) shows how a perpendicular magnetic field moves the hysteresis, associated with the CDW phase transition above RT. Figures 4 (c) and (d) illustrate the way a perpendicular magnetic field alters the CDW transition temperatures during successive heating and cooling sweeps. In this device, the change in the onset of hysteresis with the magnetic field was non-monotonic. Heating the devices in the vicinity of the NC-CDW – IC-CDW transition reduces the transition temperature as the magnetic field increases and then saturates once the field reaches ~*5 T*. Cooling the device around IC-CDW – NC-CDW transition first raises the temperature of the phase transition as the magnetic field increases to ~*2 T*, then decreases and finally saturates at ~*4 T*. We are aware of one published report about the effect of magnetic field on the NC-CDW – IC-CDW phase transition in bulk 1$T$-TaS$_2$[37]. The authors also observed a non-monotonic trend for both NC↔IC transitions, $T_{H,L}$, which were described with the model: $T_{H,L}(B)=T_{H,L}(0)+\alpha B+\beta B^2$. The temperature curve can be described as a competition between magnetic field localization and Balseiro–Falicov effect, where $\alpha$ is attributed to the Balseiro-Falicov model and $\beta$ is associated with the effect of localization[38,58,59]. Due to the imperfect Fermi nesting in 1$T$-TaS$_2$[60], the Fermi surface can be modified by an applied magnetic field. The field suppresses small electron-hole pockets and open a gap at the Fermi level, thereby improving the effective nesting conditions. As a result, the CDW transition temperature can initially increase up to a critical field. Beyond this point, the localization effect becomes dominant, reinforcing and stabilizing the incommensurate CDW phase and ultimately reducing the transition temperature. In our case of thin-film channels, in the NC→IC transition (Figure 4 (c)), the temperature drops steadily, while for the IC→NC transition (Figure 4 (d)), the temperature first rises, then decreases, like the bulk crystal case.

[**Figure 4: Magnetoresistance and dependence of the CDW phase transition on magnetic field.** (a) The magnetoresistance at various temperatures in the C-CDW phase, where the magnetic field is perpendicular to the channel. The data in (a) are for device 3. (b) The NC→IC and the IC→NC phase transition are shown in resistance vs temperature for different applied magnetic fields. (c)

The temperature shift of the NC→IC and (d) IC→NC phase transition due to the applied perpendicular magnetic field to the channel. The data in (b-d) are for device 4.]

Next, we consider the effect of the magnetic field on the electric field driven CDW domain depinning threshold, $V_D$, and phase transition $V_H$. The *I-V* characteristics of a device with a channel thickness of *~15 nm* at *T = 200 K* at different values of the perpendicular magnetic fields are shown in Figure 5 (a). The magnetic field changes the position of the NC-CDW—IC-CDW hysteresis, without changing its area. The change is irreversible. Once the magnetic field is removed, $V_H$ and $V_D$ maintain the same value. The magnetic field induced modulation of $V_H$ increases monotonically with field strength, in contrast to the electric gating, which becomes nonmonotonic at higher bias. The voltage modulation from the magnetic field for the depinning threshold, $V_D$, and for the onset of the hysteresis, $V_H$, are shown in Figure 5 (b) and (c), respectively. The voltage modulation of the domain depinning threshold $\Delta V_D/V_D$ increases to *~65%* at *2 T* and then saturates. This large increase can be attributed to localization sites induced by the magnetic field acting as strong pinning centers[9]. These defects significantly impeded domain rearrangement, requiring a higher electric field to initiate depinning[3,4]. The observed saturation suggests that the system reaches a state of maximal pinning to magnetically induced localization sites. Increasing the magnetic field further yields no additional pinning configurations. For the small magnetic field, $B \lesssim 0.3\ T$, the voltage modulation of the NC→IC-CDW transition, $\Delta V_H/V_H$, linearly increases to *~1%*. As the field grows further to *0.5 T,* the voltage modulation drops slightly and then linearly increases to *~4%* at *B = 9 T*. The linear monotonic increase in $V_H$ when increasing the perpendicular magnetic field is consistent with a stabilization of the NC-CDW phase via the Balseiro-Falicov mechanism[58,59]. Magnetic fields can reduce the transition temperature in 1*T*-TaS$_2$ through localization-induced pinning[37,38]. However, past the depinning threshold $V_D$, the electric field drives dynamic domain rearrangement, likely suppressing the static pinning effects of localization. With localization effects suppressed, the Balseiro-Falicov effect becomes dominant[37,38]. Here, the magnetic field improves Fermi surface nesting, linearly increasing the phase transition temperature[58,59]. This linear increase is substantial, deterministic, and nonvolatile, making 1*T*-TaS$_2$ devices promising for use in memory technologies.

[**Figure 5: Magnetic field effect on CDW depinning and phase transitions**. (a) *I-V* characteristics of a *h*-BN/1*T*-TaS$_2$ heterostructure CDW device at *200 K* for different values of magnetic field. Increasing the magnetic field causes $V_{H,L}$ to shift monotonically to a higher threshold. (b) The voltage modulation of $V_D$ and (c) $V_H$ with respect to the magnetic field. (d) Resistance vs magnetic field for the device at *6.77 V* and *200 K*. The data are for device 5.]

There is an interesting question related to the possibility of inducing the CDW phase transition in quasi-2D CDW 1*T*-TaS$_2$ with a magnetic field at RT. Figure 5 (d) shows the channel resistance as a function of the magnetic field when the device is biased near the NC→IC transition at $V_{DC}$=*6.77 V*. One can see an abrupt change in the resistance by a factor of *~2.5*, which corresponds to the transition from the NC-CDW to the IC-CDW phase. The ability to switch the phase of the CDW device with a magnetic field can be used for memory and other electronic applications. The observed effect of magnetic field switching of the CDW phase is reminiscent of the one utilized in the heat-assisted magnetic recording (HAMR) memory[61]. In HAMR, the amount of stored data on a magnetic device is increased by temporarily heating the disk material during writing, which makes it more receptive to magnetic effects and allows writing to smaller regions[62]. In our case, an application of the source-drain bias, larger than the CDW domain depinning threshold, produces local heating, which softens the domain, and the applied magnetic field switches the material from NC-CDW to IC-CDW phase. We experimented with electrical and magnetic fields were conducted with devices that had lateral dimensions of several micrometers. The C-CDW domain size of the NC-CDW phase near room temperature is *7 nm - 10 nm[63]*. Whereas the C-CDW phase at low temperatures shows coherence lengths ranging from tens to hundreds of nanometers, limited by crystal quality or stacking faults[17]. Based on our results, which show non-monotonic trends for the domain depinning and irreversible or partially reversible changes with the NC-CDW – IC-CDW phase transition, one should downscale the device structures to address individual C-CDW domains for further enhancement of 1*T*-TaS$_2$ device functionalities. While the demonstrated CDW voltage-controlled oscillators and information processing networks based on them utilized the two-terminal 1*T*-TaS$_2$ devices [26, 28, 29, 64, 65], the field-effect transistor-type structures, considered in this work, expand the functionality of such devices and networks, allowing for better control of the CDW phase transitions and domain depinning.

**Conclusions**

We have examined how perpendicular electric and magnetic fields influence CDW domain depinning and phase transitions in thin-film, quasi-2D 1$T$-TaS$_2$ using both top-gate and bottom-gate field-effect device architectures. Because the NC-CDW $\leftrightarrow$ IC-CDW transition occurs above RT, these measurements directly probe field-tunable behavior under technologically relevant conditions. Our results show that electrical gating produces a non-monotonic modulation of the depinning threshold, which stands in clear contrast to the monotonic gating effects observed in quasi-1D CDW materials. This work demonstrates that the interplay between carrier density, domain-wall connectivity, and pinning in a 2D CDW system is fundamentally different from that in 1D systems. We also find that a perpendicular magnetic field increases the depinning threshold and can induce the NC-CDW $\rightarrow$ IC-CDW phase transition, revealing a previously unreported magnetic-field route for controlling CDW phase stability in 1$T$-TaS$_2$. These observations highlight the sensitivity of the CDW electron–lattice condensate to multimodal external stimuli and provide new insights into how to influence the collective electronic behavior in quasi-2D CDW materials. In sum, these findings advance our understanding of field-coupled CDW dynamics in 1$T$-TaS$_2$ and establish electric- and magnetic-field control as viable strategies for engineering functional CDW devices. Looking forward, the greatest opportunities for enhanced functionality lie in device downscaling to enable the manipulation of *individual* CDW domains within the NC-CDW phase. Such capabilities would open pathways toward nanoscale, low-power electronic and information-processing technologies based on 2D CDW materials.

## 2. Experimental Section

### 4.1 Device Fabrication

Bulk 1$T$-TaS$_2$ single crystals were mechanically exfoliated using Nitto tape onto p-doped Si substrates with a *300 nm* SiO$_2$ gate dielectric. Thin 1$T$-TaS$_2$ films were identified by optical microscopy and immediately capped by dry transfer of hexagonal boron nitride (*h*-BN) using PDMS stamps, forming *h*-BN/1$T$-TaS$_2$/SiO$_2$/p$^+$Si heterostructures that protect the charge-density-wave (CDW) state from oxidation and environmental degradation. The source–drain electrodes were defined by electron-beam lithography, followed by selective removal of the *h*-BN over the contact regions using atomic layer etching to expose the underlying 1$T$-TaS$_2$. Metal contacts (*10*

*nm*/80 *nm* Ti/Au) were deposited by electron-beam evaporation. The resulting channels had lengths of *1 μm — 3 μm* and widths of *1 μm — 3.5 μm*. Two-terminal geometry was used for transport measurements throughout this work. Independent contact characterization showed that the extracted contact resistance was negligible compared to the channel resistance over the relevant temperature and bias range, justifying the use of a two-terminal configuration. For electrostatic gating, three-terminal devices were fabricated in which a local top-gate electrode was patterned on the *h*-BN capping layer, using the *h*-BN as the top-gate dielectric, while the p-doped Si substrate acted as a global back gate through the *300 nm* $SiO_2$. The Si back gate was contacted using silver paste after removing the $SiO_2$. For cryogenic and magneto-transport measurements, devices were wire-bonded using a Hybond 626 ball bonder with 1 mil *99.99%* Au wire onto either ceramic chip carriers compatible with a Lakeshore PPMS probe or Quantum Design ETO pucks for use in the standard resistivity and rotator probes.

## 2.2 Electrical and Magneto-Transport Measurements

Temperature-dependent transport measurements were performed in a Lakeshore TTPX cryogenic probe station, where devices on $p^+Si/SiO_2$ were contacted directly with tungsten micromanipulator probes on gold pads. The sample temperature was controlled by a Lakeshore 336 temperature controller operated via MeasureLINK software, using heating and cooling rates of *2 K/min*. Current–voltage (*I–V*) characteristics and electrostatic gating measurements (using both top gate and back gate) were acquired with an Agilent B1500A semiconductor parameter analyzer. For *I–V* measurements, the bias was swept at *200 mV/s* with a step of *1 mV*; the resulting I–V curves were numerically differentiated to obtain *dI/dV*, enabling analysis of current fluctuations and the CDW depinning threshold. Magneto transport measurements were carried out in a Quantum Design Physical Property Measurement System (PPMS) using either a standard resistivity probe or a rotator probe for angle-dependent studies. Devices wire-bonded to ceramic chip carriers (Lakeshore probe) or ETO pucks were measured using a Lakeshore M81 synchronized source–measure system to obtain resistance as a function of temperature and magnetic field, with the PPMS magnet and cryostat providing field and temperature control. The Agilent B1500A was used both in the TTPX and in the PPMS (via the Lakeshore chip-carrier probe and ETO pucks) to track the evolution of electrical hysteresis and CDW domain depinning under a magnetic field. Quasi-static *I–V* sweeps with the same sweep conditions as above were repeated at fixed temperatures while varying the magnetic field, and the corresponding field-dependent $I$–$V$ and $dI/dV$ characteristics were analyzed to quantify the hysteresis and depinning evolution.

## Acknowledgments

The work at UCLA was supported, in part, by the Vannevar Bush Faculty Fellowship (VBFF) to A.A.B. under the Office of Naval Research (ONR) contract N00014-21-1-2947. The work at the University of Georgia was supported, in part, *via* the subcontracts of the ONR project N00014-21-1-2947. The nanofabrication of the test structures was performed in the California NanoSystems Institute (CNSI).

## Conflict of Interest

The authors declare no conflict of interest.

## Author Contributions

A.A.B. conceived the idea, coordinated the project, and contributed to data analysis. J.O.B. fabricated devices, conducted measurements, and led the data analysis. M.T. assisted with device fabrication, contributed to data analysis, and assisted in writing the manuscript. N.S. synthesized and characterized bulk crystals of $1T$-TaS$_2$. T.T.S. supervised materials synthesis and contributed to material characterization. J.O.B. and A.A.B. wrote the manuscript. All authors contributed to manuscript preparation.

## The Data Availability Statement

The data that support the findings of this study are available from the corresponding author upon reasonable request.

## Supplemental Information

The supplemental information is available on the journal website free of charge.

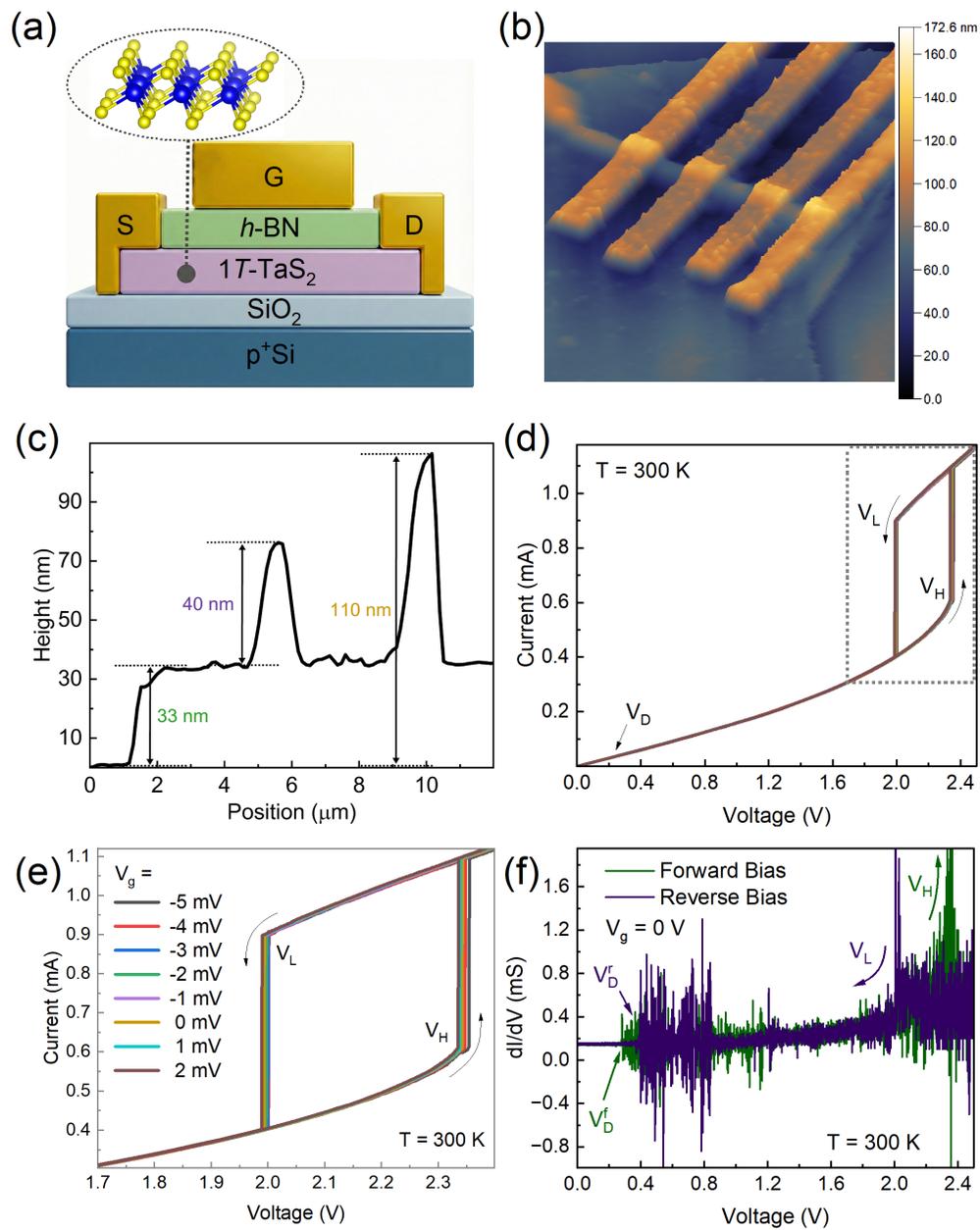

Figure 1

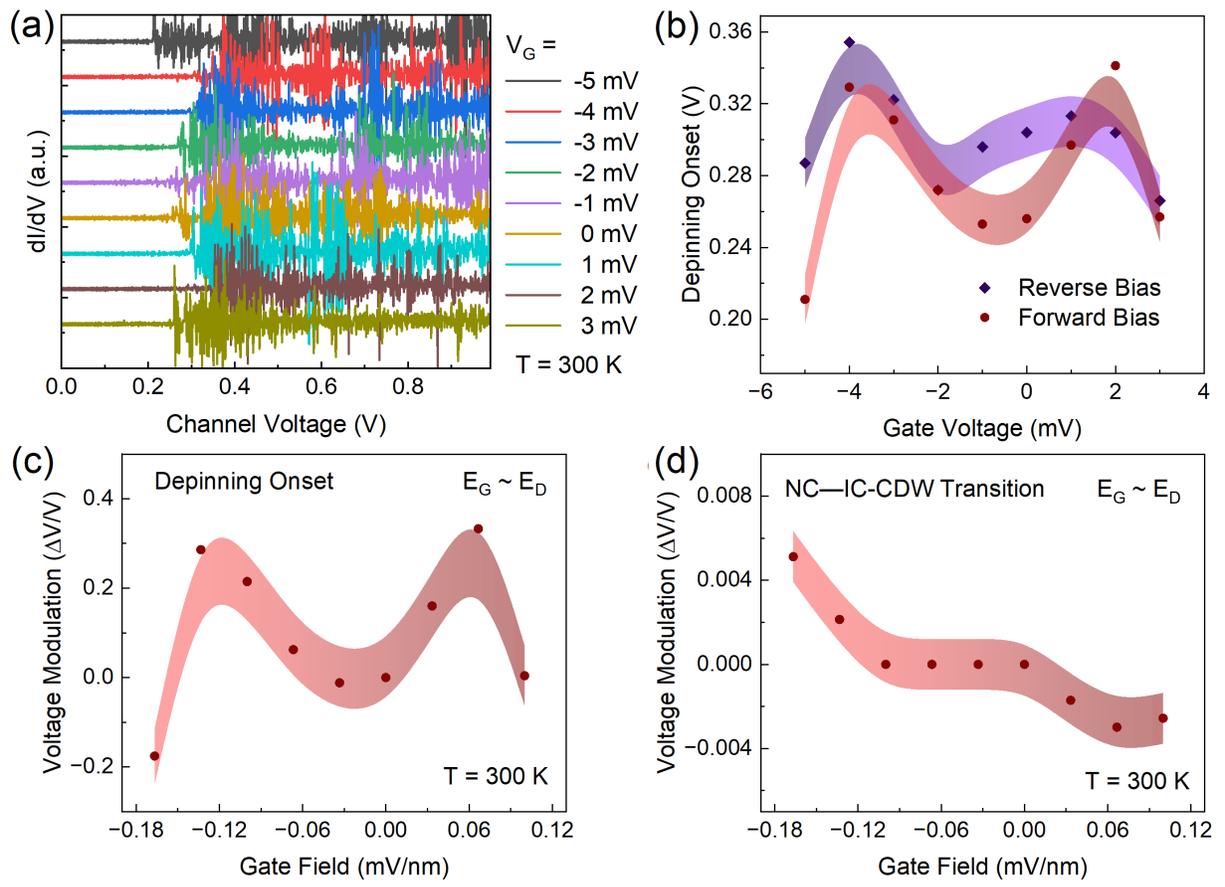

Figure 2

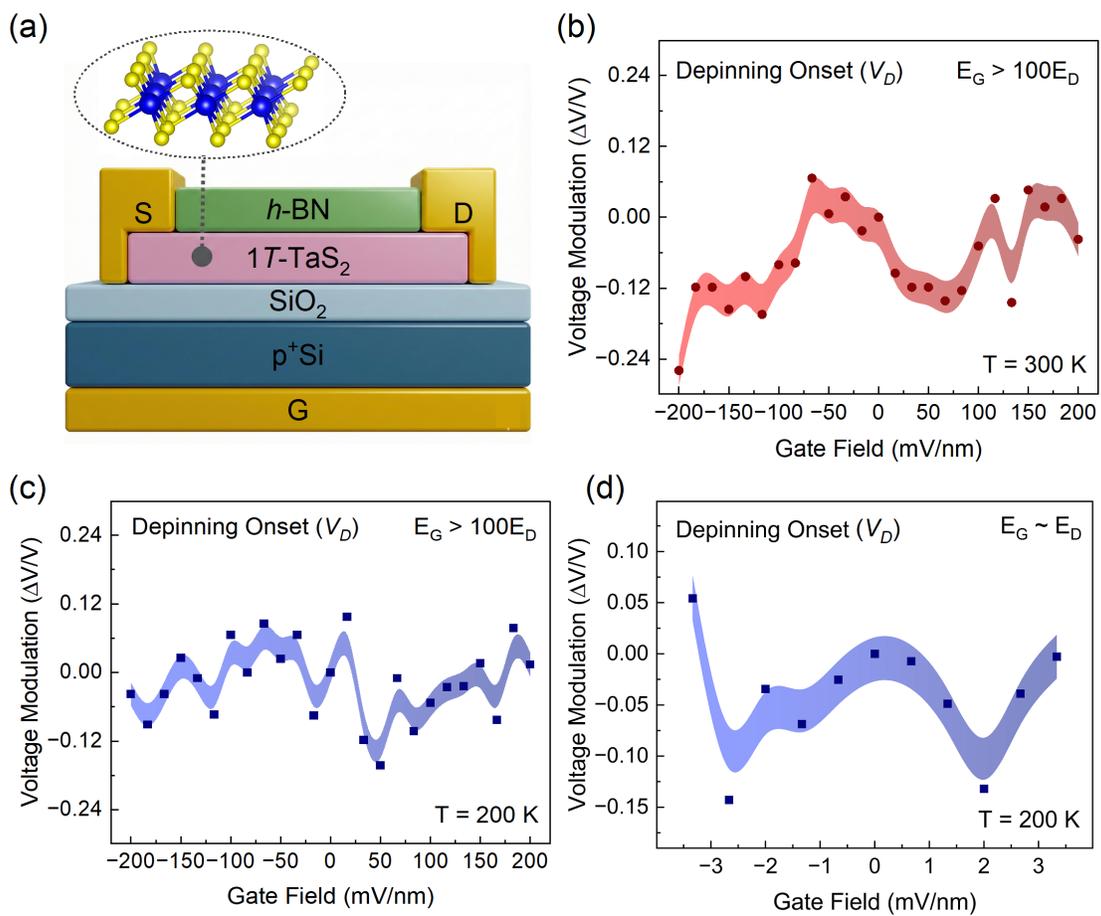

Figure 3

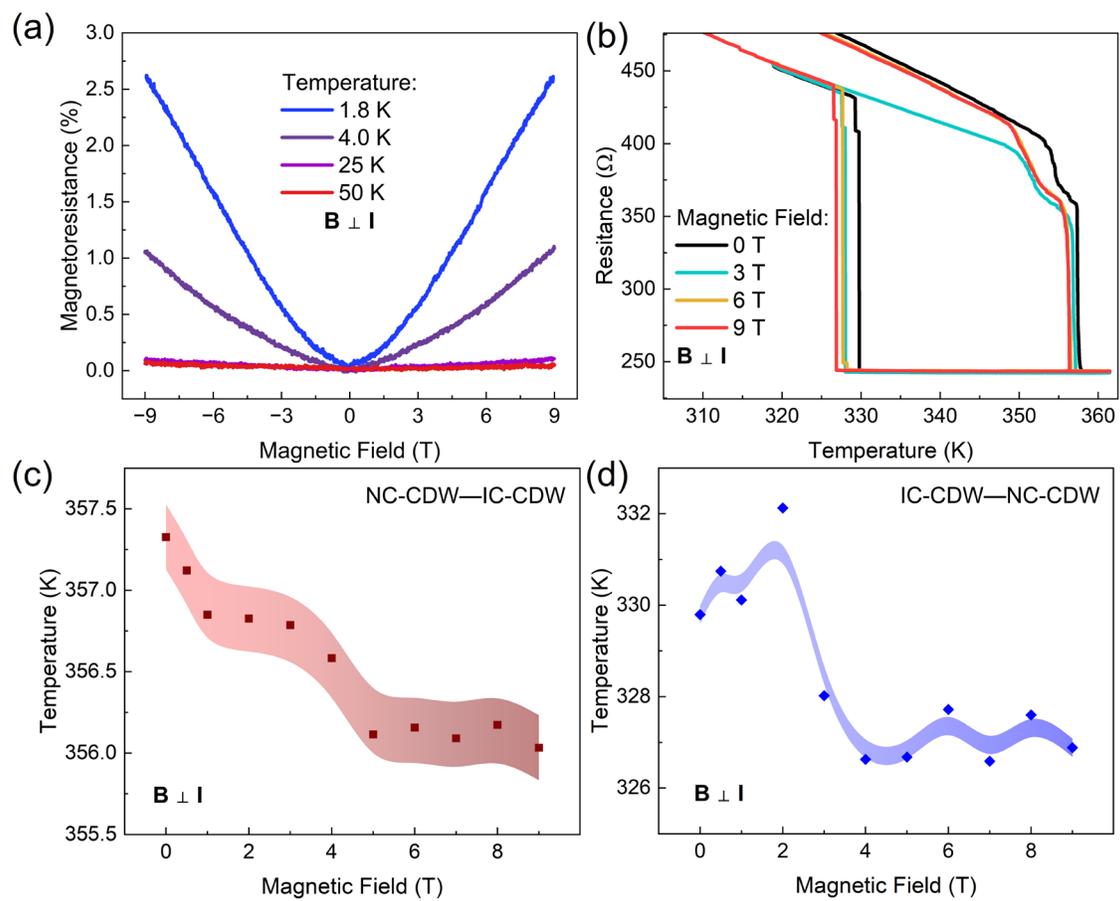

Figure 4

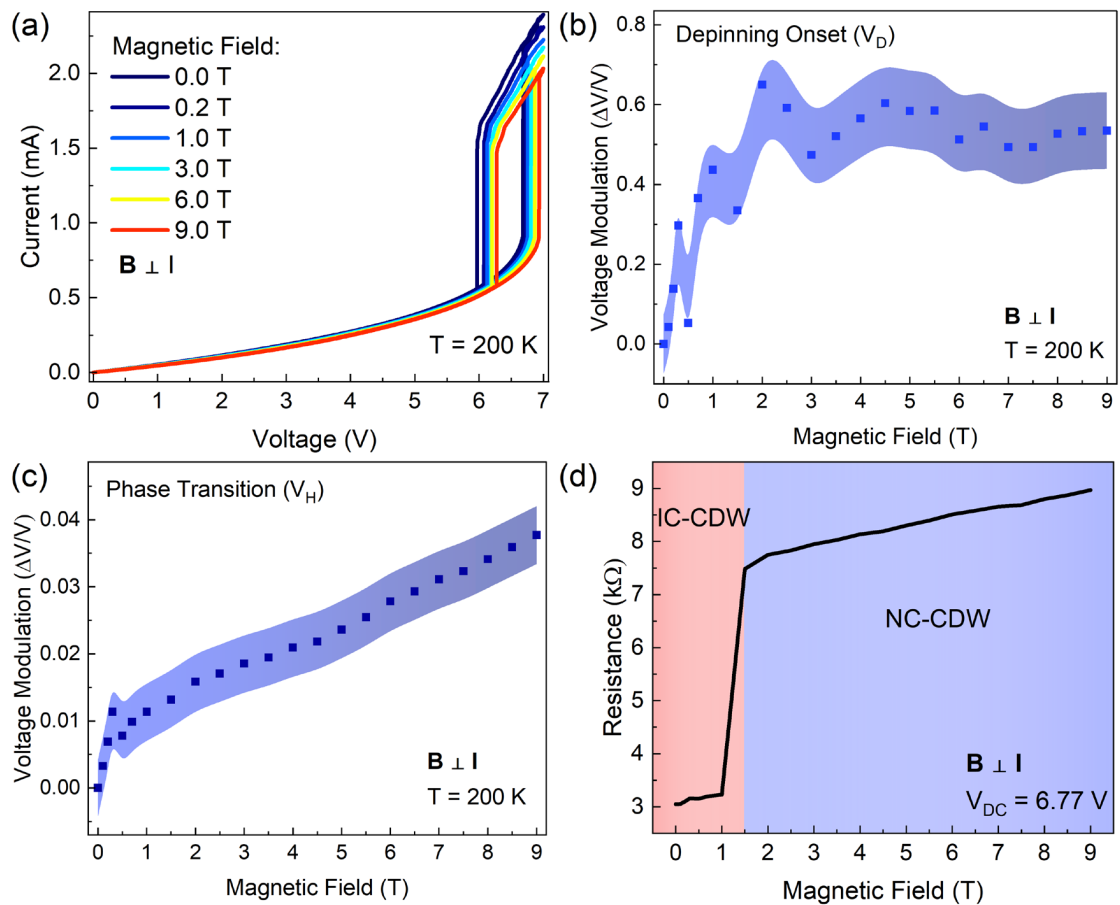

Figure 5